\documentclass[useAMS,usenatbib]{mn2e}
\usepackage{epsfig}
\title[Identifying Isolated galaxies, Pairs and Triplets]
  {High-order 3D Voronoi tessellation for identifying Isolated galaxies, Pairs and Triplets}

  \author[A. Elyiv et al.]
  { A.~Elyiv$^1$, O.~Melnyk$^{1,2}$, I.~Vavilova$^{1,3}$ \\
  $^1$Main Astronomical Observatory, Academy of Sciences of Ukraine, 27 Akademika Zabolotnoho St., 03680 Kyiv, Ukraine\\
  $^2$Astronomical Observatory, Kyiv National University, 3 Observatorna St., 04053 Kyiv, Ukraine\\
 $^3$Space Research Institute, National Space Agency of Ukraine, National Academy of Sciences of Ukraine,\\
40 Akdemika Glushkova av., 03680 Kyiv, Ukraine}
\date{Released 2008 Xxxxx XX}

\pagerange{\pageref{firstpage}--\pageref{lastpage}} \pubyear{2008}

\def\LaTeX{L\kern-.36em\raise.3ex\hbox{a}\kern-.15em
    T\kern-.1667em\lower.7ex\hbox{E}\kern-.125emX}

\begin{document}

\label{firstpage}

\maketitle

\begin{abstract}

Geometric method based on the high-order 3D Voronoi tessellation is proposed for identifying the single galaxies, pairs and triplets.  This approach allows to select small galaxy groups and isolated galaxies in different environment and find the isolated systems. The volume-limited sample of galaxies from the SDSS DR5 spectroscopic survey was used. We conclude that in such small groups as pairs and triplets the segregation by luminosity is clearly observed: galaxies in the isolated pairs and triplets are on average two times more luminous than isolated galaxies. We consider the dark matter content in different systems. The median values of mass-to-luminosity ratio are 12 $M_{\odot}/L_{\odot}$ for the isolated pairs and 44 $M_{\odot}/L_{\odot}$ for the isolated triplets; 7 (8) $M_{\odot}/L_{\odot}$ for the most compact pairs (triplets). We found also that systems in the denser environment have greater rms velocity and mass-to-luminosity ratio. 

\end{abstract}


\section{Introduction}

Physical properties of galaxies depend on the formation conditions and evolution. In addition to intrinsic evolution, galaxies are exposed to environmental influence (among others Dressler 1980, Lewis et al. 2002, Gomez et al. 2003, Einasto et al. 2003, Blanton et al. 2005, Weinmann et al. 2006, Martinez et al. 2006, Park et al. 2007, 2008). The density of galaxies (number of galaxies per volume unit) or luminosity density as well as the number of galaxies in group/cluster or distance to the nearest galaxy is often implied as regarding the environment. The influence of an environment can be found till 1 Mpc and even farther (Kauffmann et al. 2004, Blanton et al. 2005, Park et al. 2007, 2008), where small galaxy groups are observed. The study of  "enviromental effects" in such poor galaxy groups is helpful for understanding the galaxy evolution on intermediate scales between isolated galaxies and rich groups/clusters. 

The isolated galaxies that have not sufficiently undergone the influence of environment allow us to consider them as the "autonomous laboratories" for studying evolutionary processes in the galaxies. Individual properties of isolated galaxies (mass, luminosity, morphology, colour-index etc.) can be served as the standard when studying galaxies in the different environment (see, for example, Karachentseva 1973 (KIG), Prada et al. 2003, Reda et al. 2004, Stocke et al. 2004, Verley et al. 2007). For studing of galaxies' properties in different environment it is also important to define the galaxy's isolation degree which is suitable to describe by some parameter. For example, Karachentsev and Kasparova (2005) used the tidal index for each galaxy to study global properties of nearest galaxies in different environments. Verley et al. (2007) quantified the isolation degree of KIG galaxies by two parameters: local number density and tidal strength.

It is known that with the increase of galaxy systems richness from individual galaxies to clusters, their mass increases more quickly than luminosity (Karachentsev et al. 1966, Girardi et al. 2002). The dark matter in small groups seems to be distributed in the whole volume of system in the case of compact groups and to be concentrated in the halo of individual galaxies in the case of loose groups (Mulchaey et al. 2003, Melnyk \& Vavilova 2006, Da Rocha et al. 2008). At the same time the amount of dark matter in galaxy groups is not enough for standard cosmological model (Makarov \& Karachentsev 2007). 
 
As a rule, for identifying groups by different selection methods, the principle of overdensity in comparison with background is used. The richer the group population, the overdensity is more strong and therefore more likely that such a group is physically bound. Poor groups identification depends strongly on the limiting parameters of the method. These systems can be easily confused with the random physically not bound systems. For this reason, many authors prefer to study rich groups/clusters only. That is why the elaboration of the reliable methods for identifying small groups will allows us to pay more attention these structures.  

Karachentsev (1972, 1987) and Karachentseva et al. (1979, 2000) used 2D method for pairs and triplets selection taking into account the isolation degree in comparison with the foreground and the background. The environment of each pair and triplet was inspected using POSSI, POSSII and ESO/SERC plates. This method selects effectively the isolated systems, which are compact in a projection with characteristic distance between galaxies $R \sim 50 - 100$ kpc. With appearance of the large databases and surveys (LEDA, NED, SDSS, CfA2, SSRS2, DEEP2, 2dF) and information about the radial velocities of galaxies the application of three-dimensional methods of selection became generally accepted. 3D methods of group's identification use as a rule two limited parameters: projected distance between galaxies $R$ and radial velocity difference ${\Delta}V$. For example, Barton et al. (2000, 2003), Geller et al. (2006), Woods et al. (2006) selected close pairs of galaxies from CfA2 catalogue with $R <$ 50 kpc and ${\Delta}V<$1000 km~s$^{-1}$, isolation criterion was ignored. Patton et al. (2000) investigated galaxy pairs from SSRS2. They considered all pairs with $R<$100 kpc and didn't find pairs with the signs of interaction with ${\Delta}V >$ 600~km~s$^{-1}$, but most of pairs with interaction signs had $R < 20$ kpc and ${\Delta}V <$ 500 km~s$^{-1}$ (see also Patton et al. 2002, 2005, Xu et al. 2004, De Propris et al. 2007). Nikolic et al. (2004) obtained similar results for SDSS pairs. The authors found that a star-formation rate is significantly enhanced for the projected separations less then 30 kpc. Lambas et al. (2003) and Alonso et al. (2004) considered galaxy pairs with $R <$ 1 Mpc and ${\Delta}V <$ 1000 km~s$^{-1}$ from 2dF survey. They analyzed the star formation activity in the pairs as a function of both relative projected distance and relative radial velocity and found that the star formation activity in galaxy pairs is significantly enhanced over that in isolated galaxies at similar redshifts for $R <$ 25 kpc and ${\Delta}V <$ 100 km~s$^{-1}$. Soares (2007) showed that more than half of the simulated pairs with projected distance $R\leq 50$ kpc have 3D separations greater than 50 kpc. Therefore for the selection of real physical pair it is important to take into account the signs of interaction, but that is not always possible.

Triplets of galaxies are less studied than pairs. Mainly triplets have been selected from the catalogues of groups of different population (Trofimov \& Chernin, 1995). For example, Vavilova et al. (2005), Karachentseva et al. (2005), Melnyk (2006), Melnyk \& Vavilova (2006) compared the kinematic, dynamical, configurational and morphological properties of triplets from the samples formed by different methods. They showed that physical properties of galaxy groups strongly depend on selection criteria.

The main goal of our paper is to provide the uniform selection of the single galaxies, pairs and triplets from Sloan Digital Sky Survey\footnote{www.sdss.org} for later analysis of their properties: mass-to-luminosity ratio, colour indices, morphology and others. Such investigation can be helpful for study the environmental influence and dark matter content on small scales (galaxy -- pair -- triplet). To meet this goal we made not only selection of the most isolated galaxies and systems, but also single galaxies, pairs and triplets with different isolation degrees.  

Most authors mentioned above used simple selection alghorithms for pair identification: they consider all pairs with the fixed limitation parameters $R$ and ${\Delta}V$, which describe properties of galaxy pair only as a separate system, without information about their neighbours.  Contrary to this we propose another approach. We use the second-order Voronoi tessellation for galaxy pairs identification where the fundamental elements are pairs and the third-order Voronoi tessellation for galaxy triplets identification where the fundamental elements are triplets. The geometric properties of the high-order tessellations give information about relative location of neighbouring galaxies. This allows us to analyse the correlations of group properties with environment. The high-order (second or third -order) Voronoi tessellation method has not been applied earlier for groups detection, unlike the first-order Voronoi tessellation. Using the galaxies from SDSS DR5 survey with known redshifts allows us to realize the 3D approach. 

The outline of the paper is as follows. In Section 2 we present the method and parameters. The sample of galaxies is described in Section 3. In Section 4 we discuss our results and compare them with other works. Our conclusions are presented in Section 5.

\section{The method and parameters}
\begin{figure*}
\begin{tabular}{|c|c|c|}
\epsfig{file=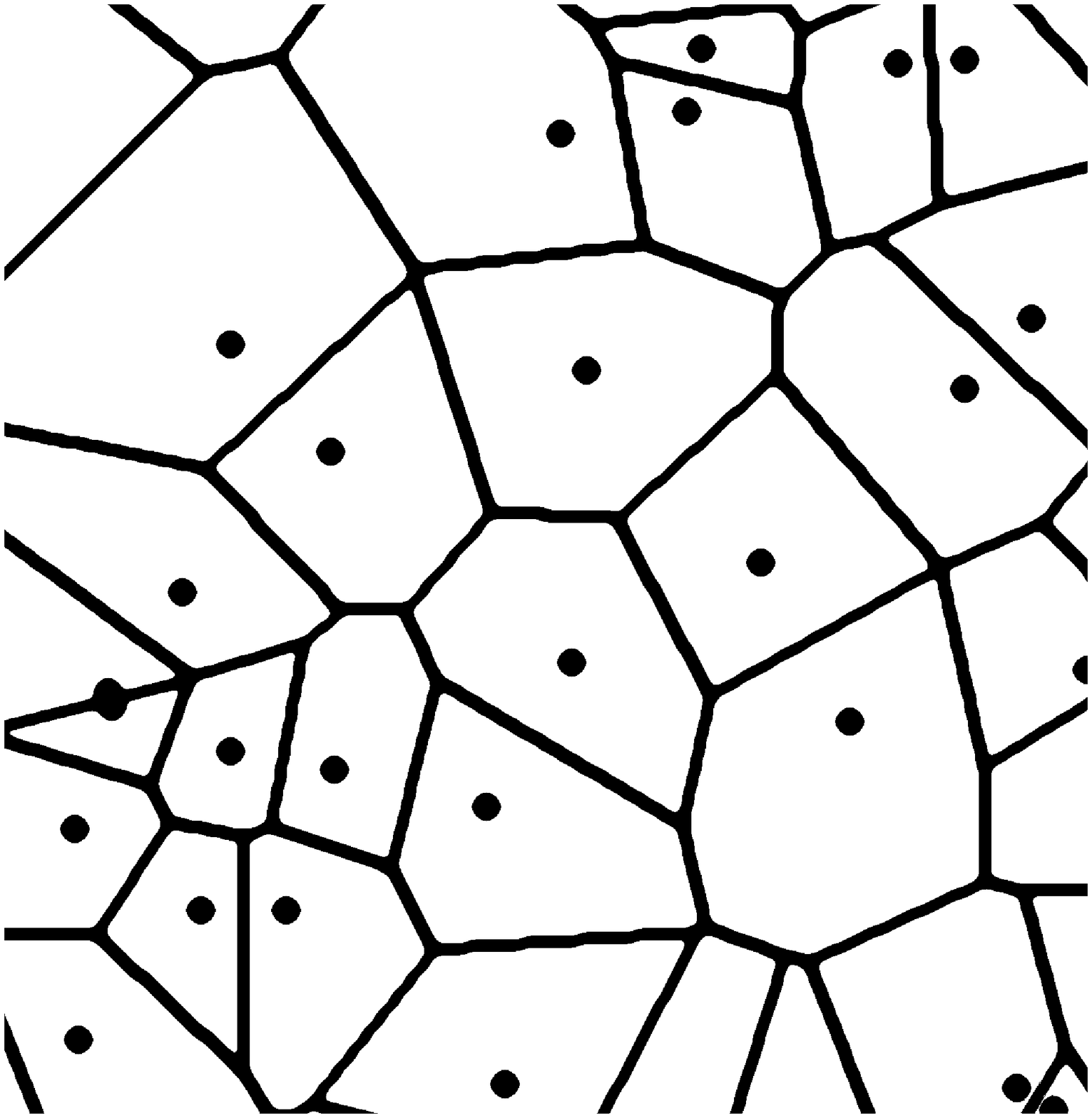,width=5.5cm} &
\epsfig{file=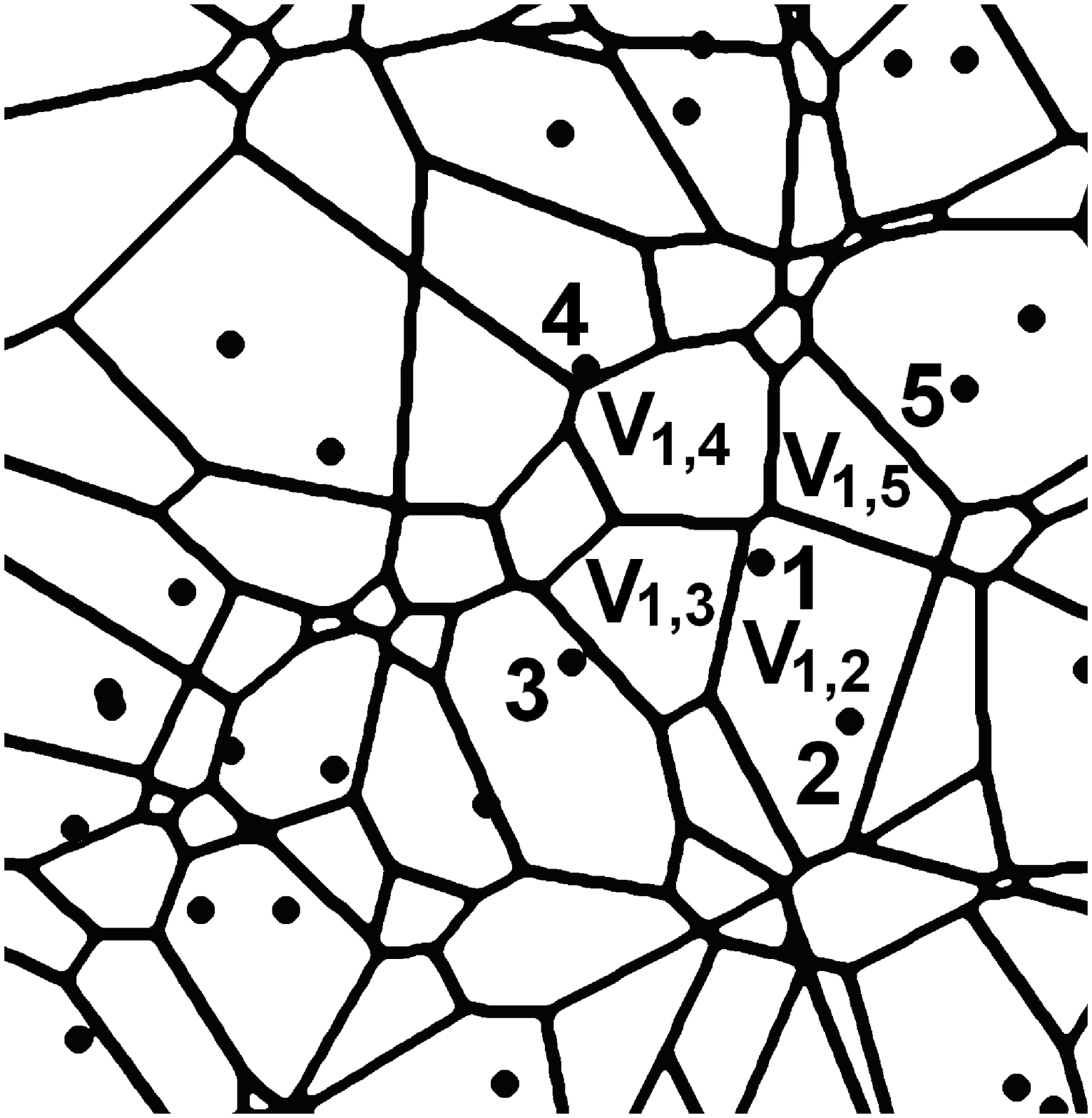,width=5.5cm} &
\epsfig{file=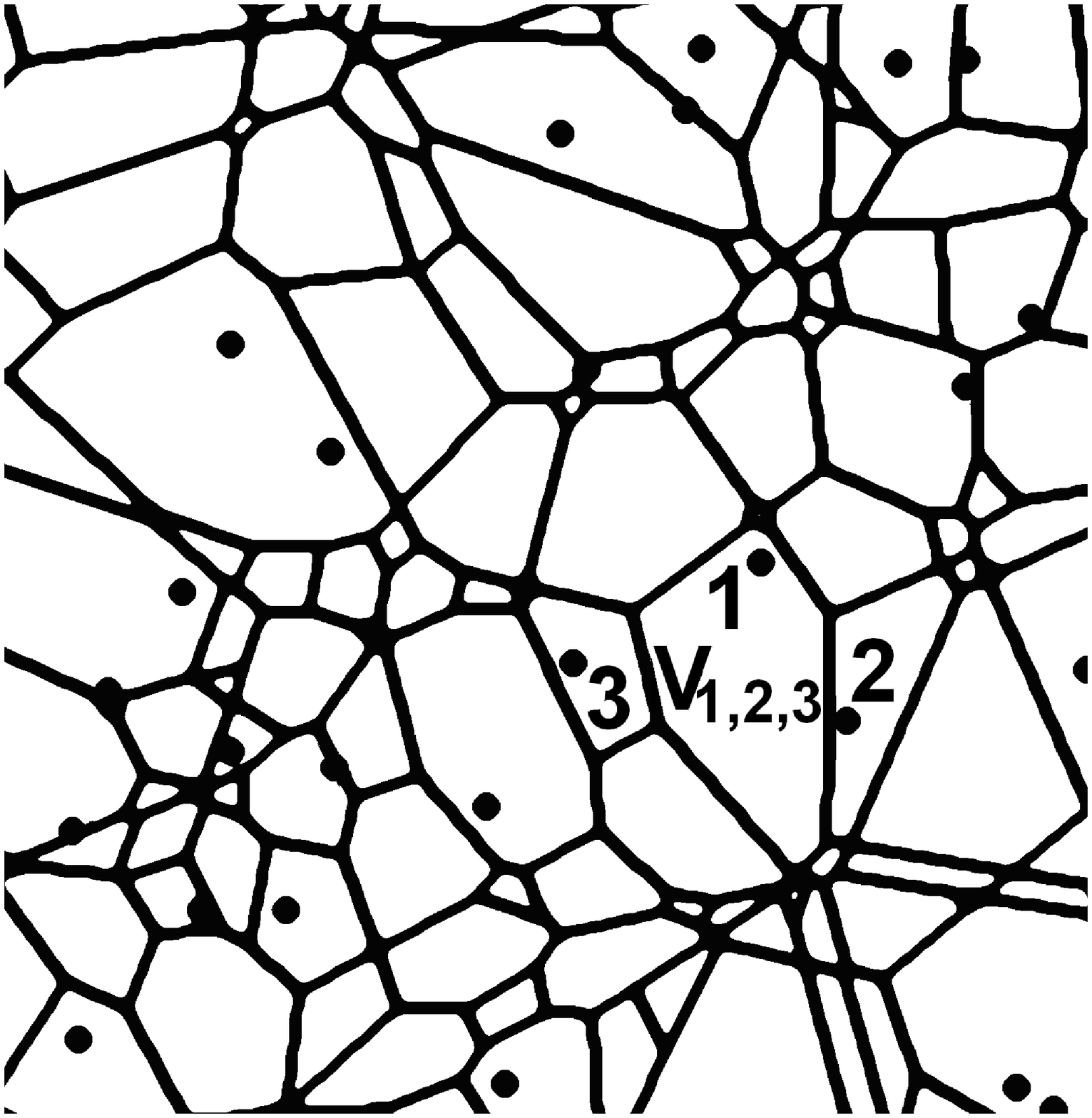,width=5.5cm} \\
a  & b & c \\
\end{tabular}
\caption{2D Voronoi tessellation of the first- a), second- b) and third- c) order for the same distribution of the random nuclei.}
\label{fig1}
\end{figure*}

Voronoi diagram was introduced by G. Voronoi in 1908. The most popular is the first-order Voronoi diagram (so called Voronoi tessellation), Fig.~\ref{fig1}a. It is a geometric method of space partition on regions. Each region consists of one nucleus and all the points of space that are closer to a given nucleus than to other nuclei (Matsuda \& Shima 1984, Lindenbergh 2002). Kiang (1966) found analytic function of Voronoi cell volume distribution for random points in 2D and 3D space. A Voronoi tessellation is used widely in astrophysics, especially for modelling the galaxy  and voids large-scale structure distribution (Icke \& van de Weygaert 1987, van de Weygaert \& Icke 1989, van de Weygaert 1994, Zaninetti  2006, van de Weygaert \& Schaap 2007), for studying periodicities in deep pencil-beam sky surveys (van de Weygaert 1991, Williams et al. 1991, Ikeuchi \& Turner 1991, SubbaRao \& Szalay 1992, Gonzales et al. 2000),  extragalactic radiosources distribution (Benn \& Wall 1995), for modeling of the cosmic microwave background anisotropy (Coles \& Barrow 1990, SubbaRao et al. 1994) etc. Ebeling \& Wiedenmann (1993) were the first who applied Voronoi tessellation for finding groups and clusters of galaxies in 2D case. Later such approach was used by Ramella et al. (2001), Kim et al. (2002), Lopes et al. (2004), Barrena et al. (2005), Panko \& Flin (2006). 3D Voronoi tessellation for galaxy groups identification was realized by Marinoni et al. (2002) and Cooper et al. (2005). The application of 3D Voronoi tessellation to  DEEP2 survey was introduced by Gerke et al. (2005). Melnyk et al. (2006) applied the first-order 3D Voronoi tessellation for the identification of galaxy groups in the Local Supercluster. The authors demonstrated that the first-order tessellation is more useful for searching the rich clusters of galaxies than the small groups. In the first-order Voronoi tesselation the key parameter is the volume of galaxy's Voronoi cell $V$. This parameter characterizes a galaxy environmental density. The condition of cluster/group membership of certain galaxy is the relatively small value of $V$. This condition is true when the galaxy is surrounded by close neighbouring galaxies. That is why the first order Voronoi tessellation is not corrected for the identifing small isolated galaxy systems (see Melnyk et al. 2006 for details). In this paper  we propose the high-order 3D Voronoi tessellation method for pairs and triplets identification. Let us consider this method below.

Contrary to the first-order tessellation (Fig.~\ref{fig1}a) the second-order tessellation for set $S$ distribution of nuclei is the partition of the space which associates a region $V_{1, 2}$ with each pairs of nuclei 1 and 2 from $S$ in such a way that all points in $V_{1, 2}$ are closer to 1 and 2 than other nuclei from $S$ (Fig.~\ref{fig1}b). Region $V_{1, 2}$ is a \textit{common cell} for nuclei 1 and 2.\footnote{However it is not necessary that these nuclei lie in the common cell. For example nuclei 1 and 5 create the common cell $V_{1, 5}$ and they do not lie in this cell.} In such a way the second-order Voronoi tessellation is available for the identification of single galaxies and pairs. 

The third-order Voronoi tessellation is appropriate for the triplets identification. It is the partition of the space which associates a region $V_{1, 2, 3}$ with each triplet of nuclei 1, 2, 3 in such a way that all points in $V_{1, 2, 3}$ are closer to nuclei 1, 2, 3 than other nuclei from $S$ (Lindenbergh 2002), Fig.~\ref{fig1}c.    

Since we work with the sample of galaxies we use galaxies as the nuclei of the Voronoi tessellation taking into account equatorial coordinates $\alpha, \delta$ and radial velocities of galaxies $V_{h}$ only. For the constructing of the 3D Voronoi tessellations it is necessary to determine the distances in 3D space. Spatial distance between two galaxies one can split on projected (tangential) distance $r$ and radial $v$ (difference of the radial velocities). We can determine the projected distance with a relatively high accuracy, while radial component has errors owing to inaccuracy of radial velocity measurement of each galaxies and existing strong peculiar velocities (due to virial motions of galaxies in groups and clusters). As a result the galaxy distribution in space of radial velocities is extended along radial component, so-called fingers-of-god effect. This is attributed to the random velocity dispersions in galaxy volume-limited sample that deviate a galaxy's velocity from pure Hubble flow, stretching out a group of galaxies in redshift space (Jackson 1972). Various authors take into account this effect in their own way in dependence on the specificity of their problem. For example Marinoni et al. (2002) chose some cylindrical window of clusterization which extended along the radial component. We propose a different way that is based on introduction of weight for a radial component (see Appendix A). Such approach allows us to avoid a problem of tangential and radial distance inequivalence and to apply high-order 3D Voronoi tessellation method.

\subsection{Second-order Voronoi tessellation. Pairs and single galaxies}

The second-order Voronoi tessellation was applied for the identification of pairs and single galaxies in the following way. Each galaxy $i$ from set $S$ forms the common cells with the certain number of neighbouring galaxies (Fig.~\ref{fig1}b). So under \textit{neighbouring galaxies} of galaxy $i$ we understand only galaxies which create common cells with this galaxy. For example, galaxy 1 creates only 4 common cells ($V_{1,2}$, $V_{1,3}$, $V_{1,4}$, $V_{1,5}$) with neighbouring galaxies 2, 3, 4 and 5, respectively. Each pair of galaxies $i, j$ is characterized by the dimensionless parameters $p_{i, j}$:

\begin{equation}
  p_{i, j}=\frac{\sqrt[D]{V_{i,j}}}{m_{i,j}},
\end{equation}
where $D$ - space dimension, $V_{i,j}$ - the area (for 2D) or volume (for 3D) of cell, $m_{i, j}$ - distance between galaxies $i$ and $j$ (see Appendix A). 

Each galaxy has the set of $p_{i, j}$ parameters. For example, galaxy 1 has the set $p_{1,2}$, $p_{1, 3}$, $p_{1, 4}$, $p_{1, 5}$. We choose the maximum value from the set of parameters for galaxy 1: $p_{max}(1)=max(p_{1,2}, p_{1, 3}, p_{1, 4}, p_{1, 5})$. In general case for galaxy $i$: $p_{max}(i)=max(p_{i,j})$ where $j$ assume numbers of $k$ neighbours. 

\begin{figure*}
\begin{tabular}{|c|c|c|}
\epsfig{file=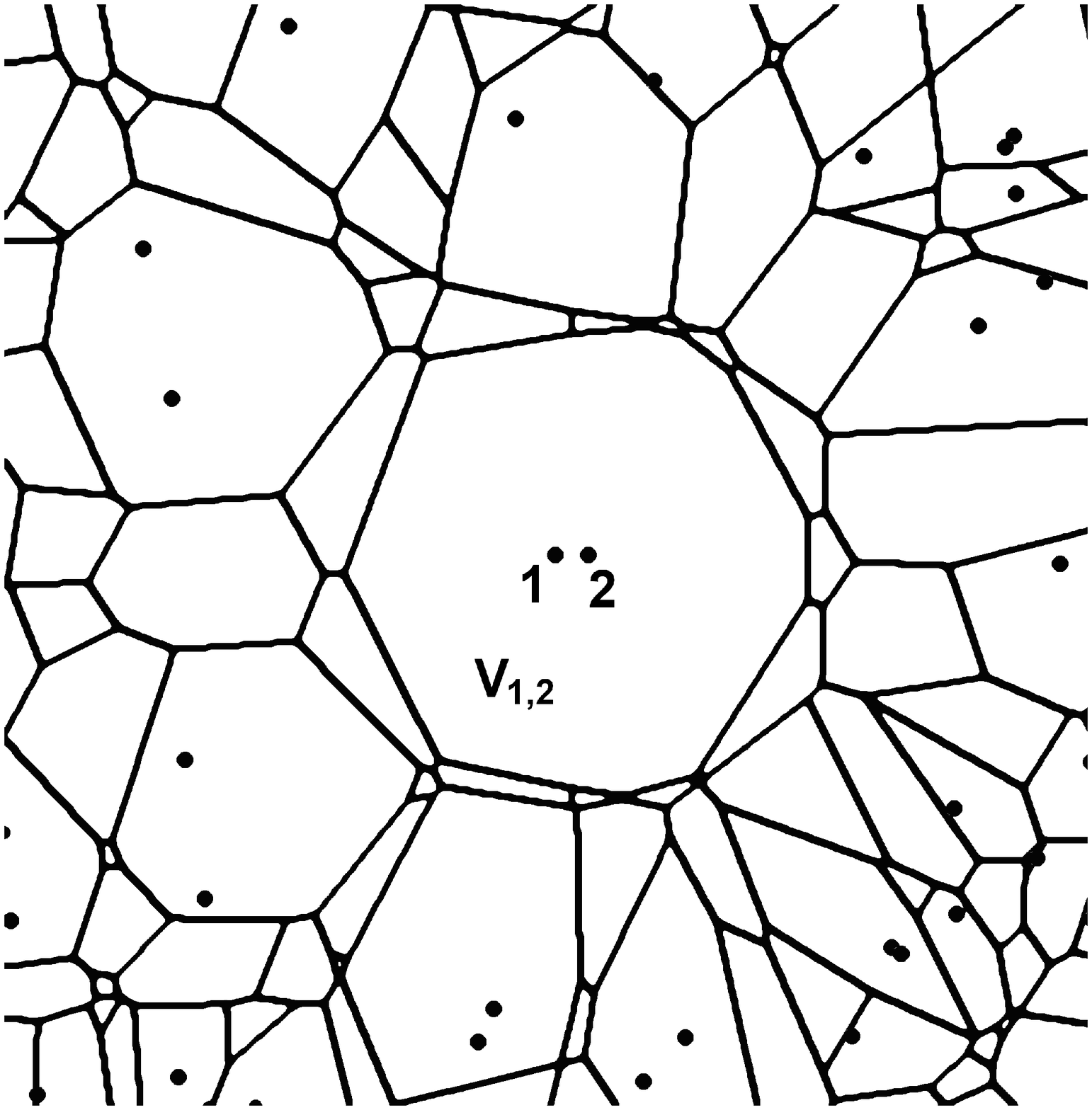,width=5.5cm} &
\epsfig{file=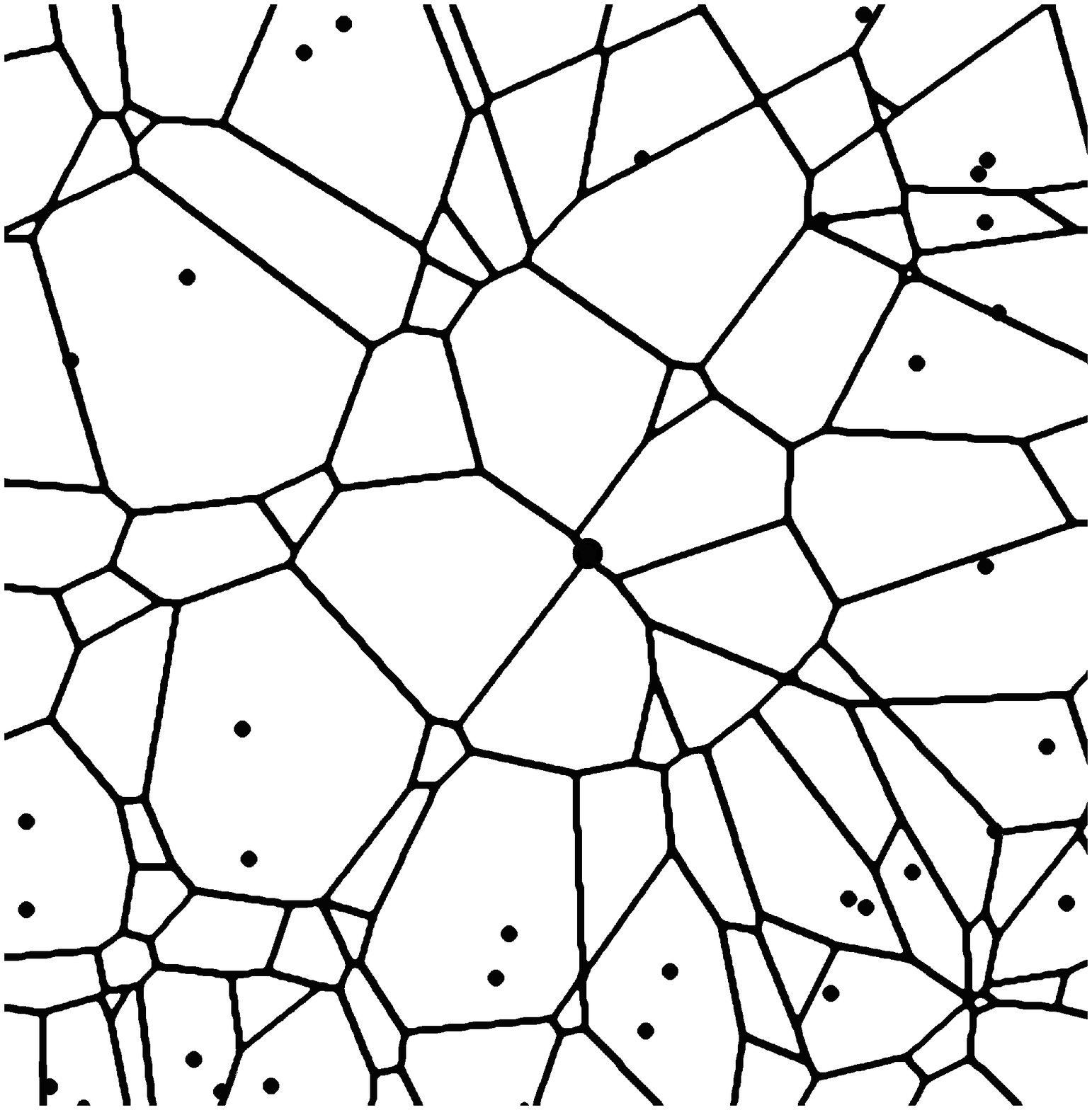,width=5.5cm} &
\epsfig{file=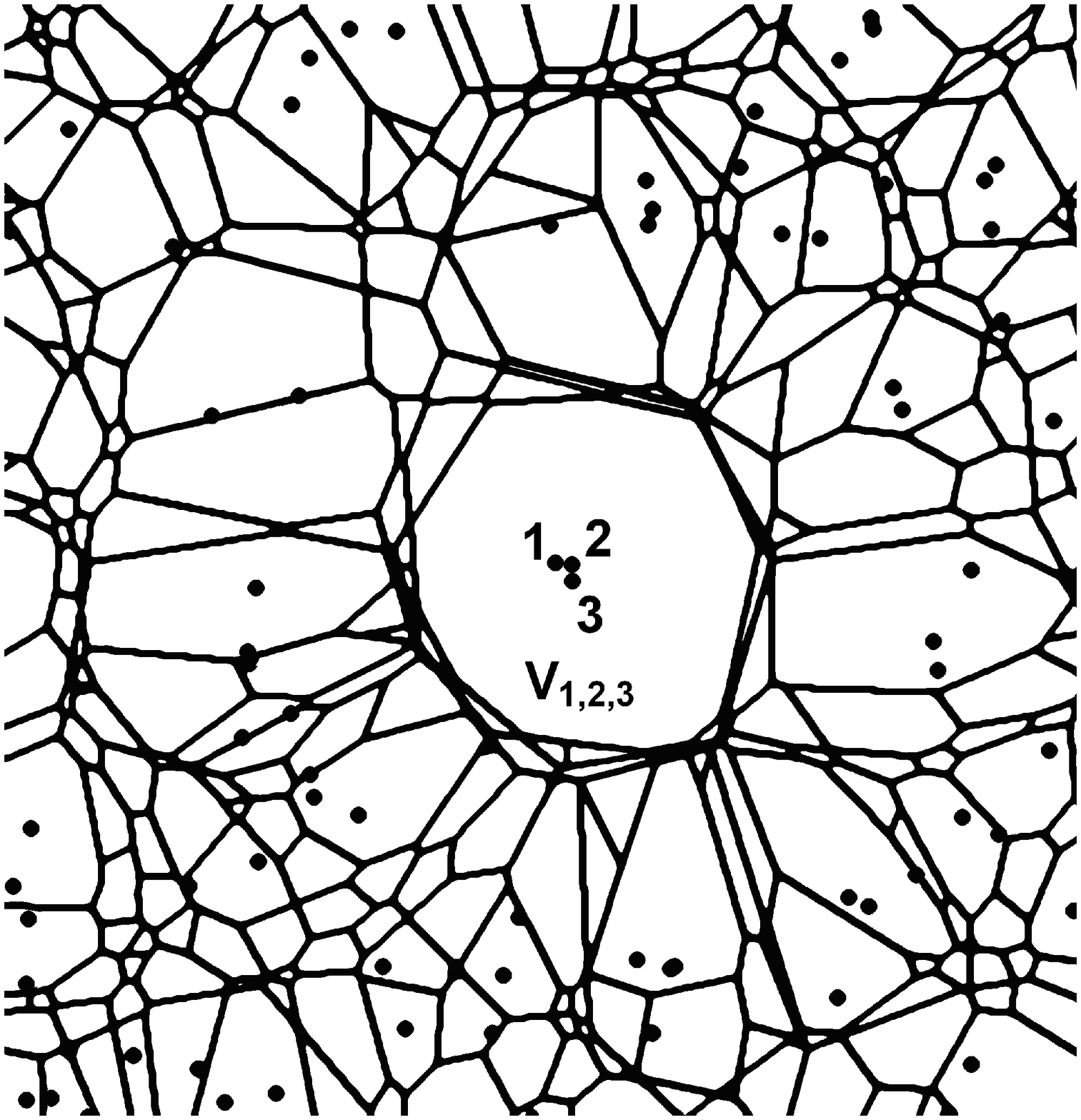,width=5.5cm} \\
a  & b & c \\
\end{tabular}
\caption{Different configurations of the galaxies: isolated close pair a) and isolated single galaxy b) in the second-order tessellation; isolated close triplet in the third-order tessellation c).}
\label{fig2}
\end{figure*}

We named the \textit{geometric pair} in the second-order Voronoi tessellation such a pair which contains two galaxies with the common cell that have the same $p_{max}$ parameters, $p_{max}(1) = p_{max}(2) = p$. The parameter $p$ characterizes a \textit{degree of geometric pair isolation}. Under degree of pair isolation we understand the overdensity under background. We used the principle which was applied by Karachentsev (1987) for the selection of isolated galaxy pairs: if all galaxies have the same angular diameter, a neighbour of a pair of galaxies should be located on the sky more than five times farther from a pair than the separation of the pair members from each other. In our 3D approach the degree of pair isolation is described by the relation between volume $V_{1, 2}$ of common cell  (it characterizes distance to neighbours) and the distance $m_{1, 2}$ between pair members according to (1). For example, strongly isolated pair has a large value of $p$ due to a large $V_{1, 2}$ and small $m_{1, 2}$ (1), see also Fig.~\ref{fig2}a. 

We also introduce the parameter $p_{e}$ which describe only \textit{pair environment} and does not depend on distance between pair members directly. We defined it as the mean value of $p_{j}(1)$ and $p_{l}(2)$ parameters of the first and second galaxy, excepting $p$ from both sets:
\begin{equation}
 p_{e}=\frac{\sum_{j=2}^{k}p_{j}(1)+\sum_{l=2}^{n}p_{l}(2)}{k+n-2},
\end{equation}
where $k$ and $n$ - number of neighbouring galaxies for 1 and 2 galaxies of geometric pair, respectively. We start sums from $j=2$ and $l=2$ for excepting $2\cdot p$, because the first galaxy is neighbour for the second galaxy and vice versa. 
Therefore $k+n$-2 is sum of neighbouring galaxies of pair members excepting of pair galaxies as neighbouring for each other.
Parameter $p_{e}$ depends on the distribution of neighbouring galaxies. Small value of $p_{e}$ points out that such pair is located in loose environment. In such case the average volume of common cells of pair components with neighbouring galaxies is relatively small, and distance between them is large, see formula (1) and Fig.~\ref{fig2}a. 

\textit{Single galaxy} is a galaxy which is not member of any geometric pair. The single galaxies are field galaxies in the environment of geometric pairs. Every single galaxy has the own neighbours, single galaxies and geometric pair members can be among them. According to the second-order Voronoi tessellation the larger is the degree of galaxy isolation, the greater is the number of neighbours (see Fig.~\ref{fig1}b in comparison with Fig.~\ref{fig2}b), but these neighbours locate farther. The best parameter which describes the isolation degree of the single galaxy  is the mean value of all parameters $p_{j}$ of this galaxy:
\begin{equation}
 s=\frac{\sum_{j=1}^{k}p_{j} }{k},
\end{equation}
where $k$ is the number of neighbours. Therefore the smaller is $s$ value, the more isolated is the single galaxy.

\subsection{Third-order Voronoi tessellation. Triplets}

Method of the third-order Voronoi tessellation can be introduced the same as the second-order approach (Fig.~\ref{fig1}c). All points of the common triplet's cell are the most closer to galaxies of this triplet than to other galaxies. Similarly to the parameter $p_{i, j}$ for pairs we set up the parameter $t_{i, j, u}$ for triplets: 
\begin{equation}
     t_{i, j, u}=\frac{\sqrt[D]{V_{i,j,u}}}{max(m_{i,j},m_{i,u},m_{j,u})},
\end{equation}
where $D$ - space dimension, $V_{i, j, u}$ - the area (for 2D) or volume (for 3D) of cell, $m_{i,j}$, $m_{i,u}$, $m_{j,u}$ - distances between galaxies in the triplet.

\textit{Geometric triplet} in the third-order Voronoi tessellation contains three galaxies that have the common cell and the same maximal parameters $t_{max}(1)=t_{max}(2)=t_{max}(3)=t$. The parameter $t$ characterizes a \textit{degree of geometric triplet isolation}. We defined parameter of \textit{triplet environment} $t_{e}$ as the mean value of parameters $t_{i}(1)$, $t_{j}(2)$ and $t_{u}(3)$, except $t$ from three sets:
\begin{equation}
 t_{e}=\frac{\sum_{i=2}^{k}t_{i}(1)+\sum_{j=2}^{n}t_{j}(2)+\sum_{u=2}^{q}t_{u}(3)}{k+n+q-3},
\end{equation}
here in case of third-order Voronoi tessellation $k$, $n$ and $q$ denote the number of \textit{neighbouring triplets} which contain galaxies 1, 2 and 3, respectively. Therefore $k+n+q-$3 is number of neighbouring triplets for certain triplet, which contain at least one galaxy from this triplet.  

It can be seen from Fig.~\ref{fig2}c and (4), (5) that for the triplet with highest degree of standing out against a background, the isolation parameter $t$ has the highest value. At the same time, if the triplet neighbours locate far from it, parameter $t_{e}$ has the small value. 

Parameters $p$, $s$, $t$ are the basic ones and define the isolation degree of galaxy pair, single galaxy or triplet  in comparison with background, respectively. Parameters $p_{e}$ and $t_{e}$ are additional ones and contain information about the distribution of the neighbouring galaxies (environment).

Similarly to the second- and third-order Voronoi tessellation it is possible to apply more high-order Voronoi tessellation for the identification of galaxy quartets, quintets and so on. 

\section{The sample}

For our investigation we used Northern part of the SDSS DR5  spectroscopic survey. Our sample is volume-limited and consist of objects that classified as galaxies. The primary sample had contained approximately 11000 galaxies with radial velocities from 2500 km~s$^{-1}$ to 10000 km~s$^{-1}$, $H_{0}$ = 75 km~s$^{-1}$ Mpc$^{-1}$. It is known that compliteness of SDSS is poor for the bright galaxies because of spectroscopic selection criteria and the diffuculty of obtaining correct photometry for object with large angular size. We tried to decrease of this effect's influence due to limiting of our sample $V_{h} > $ 2500 km~s$^{-1}$, i.e. in the way not to take into account nearest objects with the large angular diameter. Such a volume limiting also helps us  to avoid influence of Virgo cluster where strong peculiar motion exists.  We checked additionally all pairs of galaxies with a small angular resolution and excluded identical objects (parts of galaxies), which are presented twice and more in SDSS survey. All galaxy velocities $V_{h}$ were corrected for the Local Group centroid $V_{LG}$ accordingly to Karachentsev \& Makarov (1996). When we had applied the high-order Voronoi tessellation method to SDSS catalogue we limited our sample 3000 km~s$^{-1}$ $\leq V_{LG} \leq$ 9500 km~s$^{-1}$. We did not consider also galaxies that located within 2\raisebox{1ex}{\scriptsize o} near borders, because the correct estimation of Voronoi cell volume is not possible in this case. Our volume-limited sample is complete up to $17.7^{m}$ but contains also about 100 more fainter galaxies. Final number of galaxies in the sample is 6786. 

\section{The results}

We applied the second-order 3D Voronoi tessellation to our sample 6786 galaxies and obtained 2196 geometric pairs and 2394 single galaxies (65\% galaxies of whole sample are in pairs and 35\% are singles). We divided our samples of geometric pairs and singles on four equal parts by the parameter of isolation $p$ and $s$, respectively. A quater of each sample with the highest isolation degree we called as \textit{isolated} (549 pairs and 598 singles). It means that the isolated pairs and singles have $p > Q_{3}$ and $s < Q_{1}$, respectively, $Q_{3}$ is third quartile and $Q_{1}$ is the first quartile. See values of quartiles in Table~\ref{tab1d}. 

\begin{table}
\caption{Quartiles of isolation parameters.}
\label{tab1d}
\begin{tabular}{|c|c|c|c|c|}
\hline
 &	Parameters &	$Q_{1}$&$Q_{2}$&$Q_{3}$ \\
\hline
Singles & $s$ & 0.57 & 0.77 & 1.08 \\
\hline
Pairs & $p$ & 3.59 & 5.72 & 10.17\\ 	
\hline
Triplets &$t$&1.90&2.62&3.90\\
\hline
\end{tabular}
\end{table}

\begin{figure}
\epsfig{file=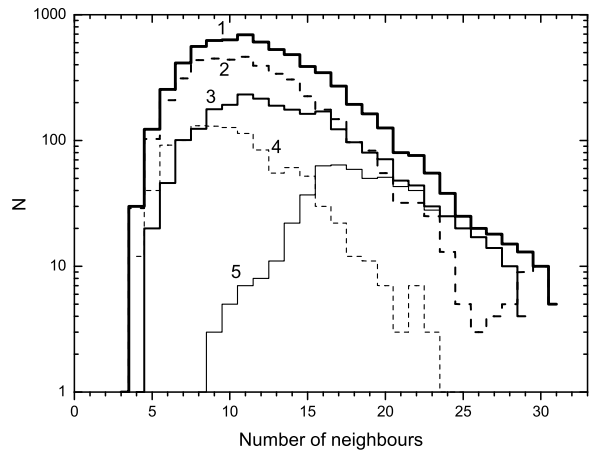,width=8.7cm}
\caption{Distributions of neighbouring galaxies number: 1) all galaxies of the sample, 2) galaxies in geometric pairs, 3) single galaxies, 4) galaxies in isolated pairs, 5) isolated galaxies.}
\label{1}
\end{figure}

Fig.~\ref{1} shows the distributions of neighbouring galaxies number for galaxies that are members of pairs and singles. We can see from Fig.~\ref{1} that number of neighbours is varried through range from 4 to 30 galaxies. Isolated galaxies have more neighbours than galaxies in other samples in average.  It is a feature of the second-order Voronoi tessellation (see above). Galaxies in isolated pairs have approximately  less by half neighbours than isolated galaxies because of the pair's neighbours distribute among  two members.

\begin{figure}
\epsfig{file=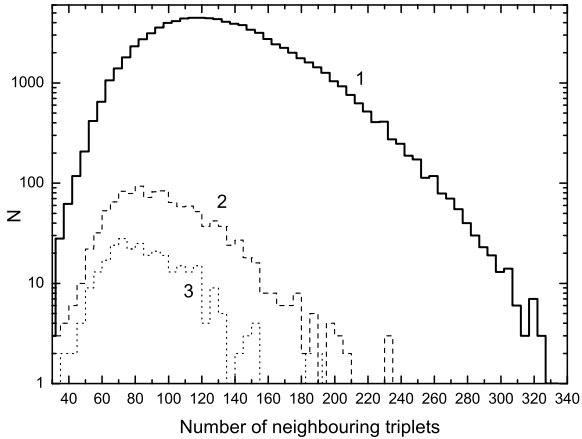,width=8.7cm}
\caption{Distributions of neighbouring triplets number: 1) all triplets of the sample, 2) geometric triplets, 3) isolated triplets.}
\label{t1}
\end{figure}

Independently we applyed the  third-order 3D Voronoi tessellation to our galaxy sample and obtained 1182 geometric triplets (52\% of whole sample). The quater (297) of triplet sample with the highest isolation degree $t>Q_{3}$ we called as \textit{isolated}. Values of all quartiles for triplets can be found in Table~\ref{tab1d}. The distribution of number of neighbouring triplets is drawn in Fig.~\ref{t1}. As can be seen, this picture looks the same as for distribution of neighbouring galaxies number in case of the second-order Voronoi tessellation (Fig.~\ref{1}).

\subsection{Comparison with other samples}

We cross-correlated our results with other samples. In the first order we compared pairs and triplets of our sample with Tago et al. (2008) groups selected by modified friends-of-friends method using the same release of SDSS. From 965 galaxies of Tago's pairs that are located in our region of investigation, 686 galaxies (343 pairs or 71\%) coincide with our pairs with different isolation degree. Median value of isolation parameter $p$ is 9.89 that does not differ strongly from our isolation limit ($Q_{3}$=10.17). Other 121 galaxies coincide with one of our pair members, median value of $p$ for these pairs is 3.94. We can conclude that they locate in the field. The remaining 158 galaxies coincide with our single galaxies having isolation degree $s$=0.87 i.e. the also associated with the field galaxies (our isolation limit is less than $Q_{1}=0.57$). 63 triplets (51\%) of Tago et al. (2008) triplets that fall in our region coincide with our triplets. Median value of isolation degree for these triplets $t$ = 4.83 is higher than our isolation limit $Q_{3}$=3.90. 

We also compared our pairs with 28 isolated Karachentsev's (1987) pairs that located in our region of investigation: 16 pairs coincide with our pairs (median value of isolation degree $p$ = 21.16 is very high). Among 7 our "pairs" one of the component is interacting galaxy which corresponds to Karachentsev's pair, second component is more fainter galaxy, these "pairs" are not isolated (median value of $p$ is 3.95). Such cases appear because of difficulties in spectroscopic measurement of interacting (very close) galaxies in SDSS. The remaining 5 pairs coincide by components with different pairs and single galaxies with small isolation degree because of these pairs are surrounded by fainter satellites. We found that 10 from 36 KIG galaxies coincided with our single galaxies, but 24 KIG galaxies fall in our pairs with different isolation degree (median value is not very high $p$=5.8). Such difference can be explained by the presence of fainter galaxy in immediate vicinity of KIG galaxy. Actually, all 34  isolated galaxies with fainter satellites from paper Prada et al. (2003), which locate in our region, coincide with our pairs and triplets. 

We can conclude: our singles, pairs and triplets of galaxies obtained by the high-order Voronoi tessellation are in a good agreement with other samples, especially it concerns the systems with the high isolation degrees. Some lack of coincidence can be explained by differences in primary catalogues, for example, by the magnitude depth or spectroscopy of interacting galaxies in SDSS.

\subsection{Isolation and main parameters}

In addition to the isolation and environmental parameters $p$, $p_{e}$, $s$, $t$ and $t_{e}$, we calculated also other characteristics (Karachentsev et al. 1989, Vavilova et al. 2005):  root-mean-square (rms) velocity of galaxies with respect to the group centre (in km $s^{-1}$), $S_{v}=\left[\frac{1}{N} \sum_{k=1}^N(V_{LG}^k-\langle V_{LG}\rangle)^{2} \right]^{1/2}$, $S_{v} = {\Delta}V/2$ for pairs, $N$ is the number of galaxies in the group; harmonic mean radii of the system (in kpc) - $R_{h}=\left[\frac{1}{3} \sum_{i,k} R_{ik}^{-1} \right]^{-1}$, where $R_{ik} = X_{i,k}\langle V_{LG}\rangle H_{0}^{-1}$ and $X_{ik}$ is the relative angular distance; $R$ - maximal distance between galaxies in triplets, for galaxy pair $R_{h} = R$;  dimensionless crossing time 

\begin{equation}
\tau =2H_{0} R_{h} /S_{v}     
\end{equation}
expressed in units of the cosmological time $H_{0}^{-1}$; virial mass 

\begin{equation}
M_{vir} = 3\pi N(N-1)^{-1}G^{-1}S_{v}^{2}R_{h}      
\end{equation}
in $M_{\odot}$; galaxy luminosity $L_{r}$, which corresponds to Petrosian magnitude in r-band, $L_{\odot}$; mass-to-luminosity ratio $M_{vir}/L = M_{vir}/\sum L_{r}$ in  $M_{\odot}/L_{\odot}$.

\begin{table}
\caption{Physical properties of pairs and triplets in dependence on isolation parameters.}
\label{tab1}
\begin{tabular}{|c|c|c|c|}
\hline
Pairs &	$N$ &	$S_{v}$ &	$R_{h}$ \\
\hline
\multicolumn{4}{|c|}{Pairs} \\ 
\hline
All geometric & 2196 & $22^{+18}_{-11}$ & $354^{+336}_{-199}$  \\ 	
\hline
$p > Q_{1}$&	1677 &	$20^{+15}_{-11}$ & $252^{+217}_{-130}$  \\ 	
\hline
$p > Q_{2}$ &	1038 &	$18^{+12}_{-10}$ & $182^{+156}_{-91} $  \\    	
\hline
$p > Q_{3}$ & 549 &	$16^{+8}_{-8}$ & $106^{+73}_{-47}$  \\     	
\hline
\multicolumn{4}{|c|}{Triplets} \\ 
\hline
All geometric & 1182 & $41^{+20}_{-15}$&	$390^{+333}_{-184}$	\\
\hline
$t > Q_{1}$ & 893 & $38^{+16}_{-14}$	&$330^{+216}_{-148}$	\\
\hline
$t > Q_{2}$	& 582 & $35^{+14}_{-12}$	&$275^{+194}_{-109}	$\\
\hline
$t > Q_{3}$	& 297 &$31^{+11}_{-10}$	&$212^{+132}_{-83}$	\\
\hline
\end{tabular}
\end{table}

\begin{table*}
\caption{Parameters $p$ and $p_{e}$ at fixed values $S_{v}$ and $R_{h}$ for geometric pairs.}
\label{tab2}
\begin{tabular}{|c|c|c|c|c|c|c|}
\hline
Pairs & & 
$S_{v} < 10$ &	$10 < S_{v} < 22$ &	$22<S_{v}<40$ &	$S_{v} > 40$	& All $S_{v}$	
\\
\hline
& $p$ & $27.61^{+17.57}_{-12.68}$ & $18.90^{+13.44}_{-6.65}$ & $11.84^{+5.62}_{-3.88}$ & $6.14^{+2.91}_{-1.38}$ & $14.67^{+13.09}_{-5.93}$ \\
$R_{h} < 155$ &$p_{e}$& $0.90^{+0.38}_{-0.20}$ & $0.94^{+0.55}_{-0.30}$ & $0.87^{+0.38}_{-0.24}$ & $0.63^{+0.19}_{-0.14}$ & $0.85^{+0.37}_{-0.24}$ \\
& $N$ & $149$ & $153$ & $144$ & $103$ & $549$ \\
\hline
& $p$ & $8.85^{+3.37}_{ -2.74}$ & $8.47^{+4.72}_{-2.49}$ & $7.05^{+2.57}_{-2.18}$ & $4.87^{ +1.71}_{ -1.10} $ & $7.36^{+3.18}_{-2.31}$  \\
$155 \leq  R_{h} < 354$ & $p_{e}$ & $0.83^{+0.25}_{-0.18}$ & $0.83^{+0.32}_{-0.20}$ & $0.82^{+0.19}_{-0.18}$ & $0.66^{+0.14}_{-0.12}$ & $ 0.78^{+0.23}_{-0.18}$ \\
& $N$ & $146$ & $154$ & $133$ & $116$ & $549$ \\
\hline
& $p$ & $5.45^{+1.65}_{-1.43} $ & $5.23^{+2.34}_{-1.39}$ & $4.73^{+1.79}_{-1.18}$ & $4.03^{+1.31}_{-0.93}$  & $4.84^{+1.84}_{-1.21}$  \\
$354\leq R_{h} < 690$ & $p_{e}$ & $0.64^{+0.17}_{-0.13}$ & $0.63^{+0.18}_{-0.10}$ & $0.69^{+0.14}_{-0.12}$ & $0.59^{+0.15}_{-0.13}$ & $0.64^{ +0.15}_{-0.13}$ \\
& $N$ & $153$ & $130$ & $139$ & $127$ & $549$  \\
\hline
& $p$ & $3.12^{+1.32}_{-0.82} $ & $3.27^{+1.21}_{ -0.81}$ & $3.05^{+1.17}_{ -0.70}$ & $2.42^{+0.90}_{-0.43} $ & $2.83^{+1.10}_{ -0.65}$ \\
$R_{h}\geq 690$ & $p_{e}$ & $0.47^{+0.11}_{-0.09}$ & $0.46^{+0.09}_{-0.09}$ & $0.52^{+0.07}_{ -0.10}$ & $0.47^{+0.10}_{ -0.08}$ & $0.48^{+0.10}_{-0.09}$ \\ 
& $N$ & $101$ & $112$ & $133$ & $203$ & $549$  \\ 
\hline
& $p$ & $7.27^{+6.74}_{-2.78}$ & $7.57^{+6.47}_{-3.32}$ & $5.63^{+3.77}_{-1.94}$ & $3.81^{+1.92}_{-1.24} $ & $5.72^{+4.45}_{-2.13}$ \\ 
All $R_{h}$ & $p_{e}$ & $ 0.71^{+0.25}_{-0.20}$ & $0.70^{+0.29}_{-0.19}$  & $0.69^{+0.26}_{-0.16}$ & $ 0.55^{+0.15}_{ -0.12}$ & $0.65^{+0.24}_{-0.15}$  \\ 
& $N$ & $549$ & $549$  & $549$ & $549$ & $2196$  \\ 
\hline
\end{tabular}
\end{table*}

\begin{table*}
\caption{Parameters $t$ and $t_{e}$ at fixed values $S_{v}$ and $R_{h}$ for geometric triplets.}
\label{tab3}
\begin{tabular}{|c|c|c|c|c|c|c|}
\hline
Triplets & & 
$S_{v} < 26$ & $26 < S_{v} < 41$ &	$41<S_{v}<61$ & $S_{v} > 61$ & All $S_{v}$	
\\
\hline
&$t$&$5.22^{+2.24}_{-2.12}$&$4.33^{+2.84}_{-1.42}$&$3.47^{+1.40}_{-1.11}$&$2.25^{+0.71}_{-0.29}$&$3.75^{+2.53}_{-1.23}$\\
$R_{h}<205$&$t_{e}$&$0.43^{+0.14}_{-0.08}$&$0.41^{+0.09}_{-0.06}$&$0.36^{+0.05}_{-0.06}$&$0.29^{+0.05}_{-0.05}$&$0.38^{+0.10}_{-0.07}$\\
&$N$&$91$&$92$&$70$&$43$&$296$\\
\hline
&$t$&$3.44^{+1.74}_{-0.82}$&$3.36^{+1.13}_{-1.04}$&$2.84^{+1.05}_{-1.64}$&$2.25^{+0.69}_{-0.29}$&$3.11^{+1.17}_{-0.85}$\\
$205\leq R_{h}<390$&$t_{e}$&$0.41^{+0.05}_{-0.07}$&$0.38^{+0.05}_{-0.07}$&$0.37^{+0.04}_{-0.06}$&$0.32^{+0.03}_{-0.04}$&$0.37^{+0.06}_{-0.06}$\\
&$N$&$81$&$87$&$83$&$44$&$295$\\
\hline
&$t$&$2.96^{+0.83}_{-0.78}$&$2.47^{+0.92}_{-0.57}$&$2.50^{+0.74}_{-0.47}$&$2.03^{+0.38}_{-0.44}$&$2.42^{+0.79}_{-0.54}$\\
$390\leq R_{h}<723$&$t_{e}$&$0.31^{+0.09}_{-0.04}$&$0.32^{+0.05}_{-0.04}$&$0.33^{+0.06}_{-0.04}$&$0.29^{+0.05}_{-0.03}$&$0.32^{+0.06}_{-0.05}$\\
&$N$&$71$&$65$&$76$&$83$&$295$\\
\hline
&$t$&$1.92^{+0.92}_{-0.36}$&$1.97^{+0.33}_{-0.25}$&$1.89^{+0.61}_{-0.35}$&$1.74^{+0.69}_{-0.31}$&$1.88^{+0.64}_{-0.38}$\\
$R_{h}\geq 723$&$t_{e}$&$0.25^{+0.04}_{-0.05}$&$0.27^{+0.03}_{-0.05}$&$0.27^{+0.04}_{-0.04}$&$0.24^{+0.02}_{-0.03}$&$0.25^{+0.03}_{-0.04}$\\
&$N$&$53$&$51$&$66$&$126$&$296$\\
\hline
&$t$&$3.23^{+1.96}_{-0.97}$&$3.08^{+1.37}_{-1.01}$&$2.65^{+1.02}_{-0.65}$&$2.03^{+0.57}_{-0.44}$&$2.62^{+1.28}_{-0.71}$\\
All $R_{h}$&$t_{e}$&$0.37^{+0.09}_{-0.09}$&$0.35^{+0.08}_{-0.07}$&$0.33^{+0.07}_{-0.05}$&$0.27^{+0.05}_{-0.04}$&$0.32^{+0.07}_{-0.06}$\\
&$N$&$296$&$295$&$295$&$296$&$1182$\\
\hline
\end{tabular}
\end{table*}

First of all we investigated how the values of rms velocity and mean harmonic radius of pairs and triplets depend on their isolation degree.  Table~\ref{tab1} presents the median values and quartiles of $S_{v}$ and $R_{h}$ for pairs and triplets in dependence on isolation parameters $p$ and $t$: all geometric pairs, pairs with $p > Q_{1}$, $p > Q_{2}$, $p > Q_{3}$; all geometric triplets, triplets with $t > Q_{1}$, $t > Q_{2}$, $t > Q_{3}$, where $Q_{1}$, $Q_{2}$, $Q_{3}$ is the first, second (median value), third quartiles, respectively. See values of $p$ and $t$ quartiles in Table~\ref{tab1d}. It can be seen from Table~\ref{tab1} that all geometric samples have largest values of  $S_{v}$ and $R_{h}$ and largest interquartile range (IQR). With the increase of isolation degree, the medians and IQR of $S_{v}$ and $R_{h}$ decrease. Pairs and triplets selected by dynamical method (Makarov $\&$ Karachentsev 2000) have medians $S_{v} = 24^{+22}_{-13}$ km s$^{-1}$, $R_{h} = 170^{+185}_{-114}$ kpc and $41^{+19}_{-18}$ km s$^{-1}$, $191^{+157}_{-88}$ kpc, respectively. The sample of triplets from Vavilova et al. (2005) has medians $S_{v}=30^{+18}_{-12}$ km s$^{-1}$ and $R_{h}=160^{+117}_{-77}$ kpc. These values better agree with samples which have $p, t > Q_{2}$ (i.e. $p>$ 5.72 and $t>$2.62 for pairs and triplets, respectively). 

We also considered how the isolation degree of pairs and triplets depends on $S_{v}$ and $R_{h}$. Table~\ref{tab2} and ~\ref{tab3} give values of parameters $p$, $p_{e}$, $t$, $t_{e}$ for pairs and triplets at the fixed values of $S_{v}$ and $R_{h}$, which correspond to intervals: $S_{v}$, $R_{h} < Q_{1}$; $Q_{1} < S_{v}$, $R_{h} < Q_{2}$; $Q_{2} < S_{v}, R_{h} < Q_{3}$; $S_{v}, R_{h} > Q_{3}$; All $S_{v}$, $R_{h}$.

Most compact pairs with $R_{h}$ less than the first quartile have the medians of $p$ by 2.5 times greater and IQR almost by three times greater than the values for sample "All $S_{v}$, $R_{h}$". That means that compact pairs are more isolated (or more contrast ones in comparison with the background) than in average galaxies in the whole sample, but also they are characterized by different isolation degree. At $R_{h} < Q_{1}$ and $p, t > Q_{3}$,  65 \% pairs (47 \% triplets) coincide.

Parameter $p_{e}$ for the compact pairs has a large dispersion also. This means that similar compact pairs are both in densest and loose environment.  Thus, the isolation degree and IQR change depending on the values of rms velocities: the most compact pairs with rms velocity less than the first quartile are most isolated. For these pairs parameter $p$ almost by 5 times greater than median of "All $R_{h}$, $S_{v}$" sample and IQR by 4.5 times greater. Most compact pairs with $S_{v} > Q_{3}$ have isolation degree the same as galaxies of whole sample.  The values of isolation parameter decrease with increasing of $R_{h}$ and $S_{v}$, thus, at $R_{h}$ greater of third quartile the $p$ becomes almost by 2 times less than the median of a whole sample. It means that these pairs are less isolated.  It is necessary to note, that $p$ significantly changes in dependence on $R_{h}$, than $S_{v}$. But all of pairs with $S_{v} >$ 40 km~s$^{-1}$ ($Q_{3}$) are characterized by small isolation degree, than pairs with $S_{v} <$ 40 km~s$^{-1}$. Additional parameter $p_{e}$ also decreases with $R_{h}$ increasing and for $R_{h} > Q_{3}$ it has the minimum value with a small variation ($p$ is also minimal). So, all wide pairs ($R_{h} > Q_{2}$) are in low dense environment. It is obvious that for the widest pairs to be isolated there are not enough free space (without galaxies). Most probably these pairs are the accidental ones in the common field. From Table~\ref{tab3} it is easy to be convinced that all above mentioned tendencies are similar for triplets but with their own values.  

\subsection{The luminosity content}

\begin{figure}
\epsfig{file=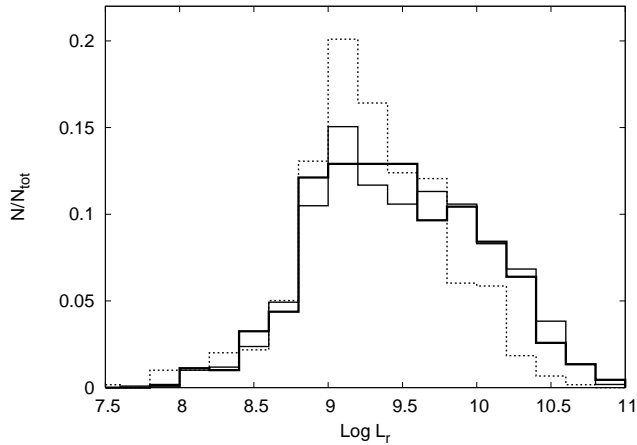,width=8.7cm}
\caption{Luminosity distributions of isolated galaxies (dotted line), galaxies in isolated pairs (solid line), galaxies in isolated triplets (thick line).}
\label{fig3}
\end{figure}

As considering the luminosity-density relation (for example Park et al. 2007, 2008 and references therein) we can note the following: galaxies are more luminous in the high density regions than in the field. Karachentseva et al. (2005) showed that galaxies even in such poor groups as the triplets are more luminous than the isolated galaxies. We compared the luminosities of single/isolated galaxies with the luminosities of galaxies in pairs and triplets. Their medians with quartiles and mean values with standard deviations are presented in Table~\ref{tab4}. Fig.~\ref{fig3} shows the distribution of galaxy luminosities. 

You can see from Table~\ref{tab4} and Fig.~\ref{fig3} that the median and mean values of luminosities are greater for galaxies in pairs and triplets than for single/isolated galaxies. The mean values of luminosities of isolated galaxies and galaxies in isolated pairs differ significantly (with the probability of $>$ 0.99 by t- criterium), but differences for "All" samples are not statistically significant. The same conclusions we made from comparison of the sample of single/isolated galaxies with galaxies in triplets. 

\begin{table}
\caption{Comparison between the luminosities of the galaxies from different samples.}
\label{tab4}
\begin{tabular}{|c|c|c|}
\hline
Sample of galaxies & \multicolumn{2}{|c|}{$L_{r}\times 10^{9} (L_{\odot})$} \\ 
\cline{2-3}
& Median & Mean (SD) \\ 
\hline
All single & $2.23^{+3.66}_{-1.15}$ &	5.17 (8.00) \\
All geometric pairs & $2.38^{+4.27}_{-1.29}$ & 5.79 (8.66) \\
All geometric triplets & $2.48^{+1.36}_{-9.14}$ & 6.00 (8.77) \\ 
Isolated & $1.80^{+3.10}_{-0.77}$ & 3.52 (4.47) \\
Isolated pairs & $2.89^{+7.20}_{-1.73}$ & 6.74 (9.46) \\
Isolated triplets & $2.68^{+6.80}_{-1.52}$ & 6.53 (9.97) \\
\hline
\end{tabular}
\end{table}

So, galaxies in isolated pairs and triplets are two times more luminous than isolated galaxies. It is necessary to note that our method of group identification doesn't  take into account individual physical characteristics of galaxies, therefore the effects of selection are absent here. The fact that for "All" samples the mean values have small difference may serve as the evidence of influence of non-physical (accidental) groups in these samples, the wide systems are dominated among them.  

\subsection{Mass-to-luminosity ratio}

\begin{figure}
\epsfig{file=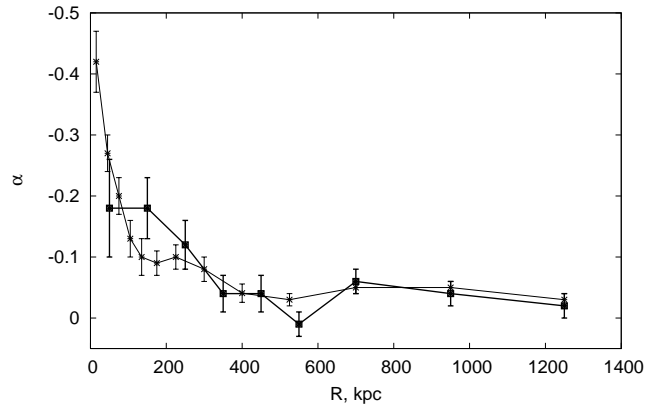,width=8.7cm}
\caption{Dependence between slope $\alpha $ and $R$: for the pairs $log(p) = \alpha \cdot log(M_{vir}/L) + b$ (thin line) and for the triplets $log(t) = \alpha \cdot log(M_{vir}/L) + b$ (thick line).}
\label{fig4}
\end{figure}

\begin{figure}
\epsfig{file=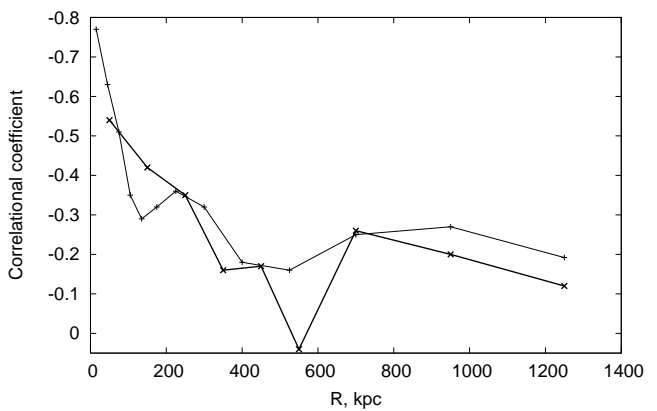,width=8.7cm}
\caption{Dependence between correlation coefficient in relations from Fig.~\ref{fig4} on $R$: for the pairs (thin line) and for the triplets (thick line).}
\label{fig5}
\end{figure}

For studying  dark matter presence and distribution in small galaxy groups we used the mass-to-luminosity ratio ($M_{vir}/L$) as the quantitative indicator of dark matter contribution. We checked the mass-to-luminosity ratio in dependence on isolation degree of pairs and triplets. We plotted the dependences of system isolation $p$, $t$ on $M_{vir}/L$ in the narrow bins of $R$, because the isolation parameters $p$, $t$ and virial mass $M_{vir}$ depend on the projected distance between the galaxies $R$ (see (1), (4) and (7)). Fig.~\ref{fig4} presents the dependence of slope $\alpha $ in $p$, $t$-$M_{vir}/L$ relation on $R$, and Fig.~\ref{fig5} presents dependence of correlation coefficient on  $R$ (details in signatures to the figures).

We see from Fig.~\ref{fig4} and 7 that if the system is located in denser environment, it has greater value of $M_{vir}/L$ ($a <$ -0.1, correlation coefficient $<$ -0.3). Moreover, such dependences are true only on the scales approximately up to 150-200 kpc for pairs and up to 250-300 kpc for triplets. For greater values of $R$ the $p$,$t- M_{vir}/L$ dependences were not observed. This conclusion only confirms our results (subsections 4.2 and 4.3) where we showed that the widest systems most probably are accidental formations, because they don't stand out against a background.  

Our results agree with the work by Park et al. (2008) where authors studied the dependence of morphology and luminosity on environment. The strong dependence of morphology on the nearest neighbour was found at the distances of about 200 $h^{-1}$ kpc. On the other hand authors (see for example Lambas et al. 2003, Alonso et al. 2004, Woods et al. 2006, Patton et al. 2000, De Propris et al. 2007) on the base of analysis of physical properties such as starformation rate and colour indices  concluded that differences between galaxies in pairs and fields galaxies are significant in pairs with $R <$20-50 kpc. Dependence on velocity dispersion is weaker, but Lambas et al. (2003) and Alonso et al. (2004) found significant differences in ${\Delta }V <$ 100 km s$^{-1}$. 

Because of maximal distance $R$ is a constant, enhancement of $M_{vir}/L$ value is a result of increase of rms velocities in groups. In other words, if the relatively compact pairs and triplets located in denser environment, then they have greater value of mass-to-luminosity ratio due to larger virial motions.  The same result was obtained by Einasto et al. (2003). The authors found that loose groups in the neighbourhood of a rich cluster are typically 2.5 times more massive and 1.6 times more luminous than groups on average, and these loose groups have velocity dispersions 1.3 times greater than groups on average. 

Table~\ref{tab5} presents the medians and quartiles of system crossing time ${\tau}$ and mass-to-luminosity ratio $M_{vir}/L$ for samples of pairs and triplets that characterized by different isolation degree and compactness. 

From Table~\ref{tab5} follows that values of $M_{vir}/L$ at $R <$ 50 kpc for pairs and $R <$ 100 kpc for triplets is 7-8 $M_{\odot}/L_{\odot }$ and increase with enhancing of $R_{h}$ faster for triplets than for pairs. 

The values of $M_{vir}/L$ are in a good agreement with the corresponding values within IQR for groups which were selected by dynamical method. The values of $M_{vir}/L$ are 18$^{+39}_{-14}$ $M_{\odot}/L_{\odot}$ and 32$^{+34}_{-22}$ $M_{\odot}/L_{\odot}$ for pairs and triplets, respectively (Makarov \& Karachentsev 2000) and for triplets from Vavilova et al. (2005) 35$^{+30}_{-23}$  $M_{\odot}/L_{\odot}$. The median of most compact triplets is in agreement with median $M_{vir}/L = 13^{+16}_{-11}$ $M_{\odot}/L_{\odot}$ obtained for interacting triplets, where star formation rate is high (Melnyk 2006). The median of isolated triplets is in a good agreement with median $M_{vir}/L = 47^{+48}_{-33}$ $M_{\odot}/L_{\odot}$ which corresponds to the sample of isolated Northern (Karachentsev et al. 1979) and Southern (Karachentseva \& Karachentsev 2000) triplets.

\begin{table*}
\caption{Medians and quartiles of the system crossing time and mass-to-luminosity ratio for pairs and triplets.}
\label{tab5}
\begin{tabular}{|c|c|c|c|c|c|c|c|}
\hline
Pairs & $N$ & $\tau $	& $M_{vir}/L$ & Triplets* & $N$ & $\tau $ & $M_{vir}/L$  \\
\hline
$p>Q_{3}$ &	519	& $1.12^{+1.24}_{-0.59}$ & $12^{+37}_{-10}$ & $t > Q_{3}$ &	297 & $1.06^{+0.88}_{-0.42}$ & $44^{+60}_{-28}$ \\
$R < 50$ & 	$133$	& $0.18^{+0.20}_{-0.07}$	& $7^{+28}_{-6}$ & $	R <100$ &	$16$ &	$0.24^{+0.07}_{-0.08}$ & $8^{+16}_{-3}$  \\
$R < 100$ &	$335$	& $0.40^{+0.47}_{-0.21}$ &	$9^{+31}_{-8}$	& $R < 200$	&70&	$0.43^{+0.49}_{-0.24}$ &	$20^{+46}_{-13}$  \\
$R < 150$ &	$549$	& $0.61^{+0.63}_{-0.32}$ &	$14^{+44}_{-11}$ & $	R < 300$&	136&	$0.70^{+0.33}_{-0.28}$ &	$24^{+41}_{-16} $ \\
$R < 200$ &	$684$	& $0.70^{+0.75}_{-0.34}$ &	$17^{+50}_{-14}$ & $	R < 400$&	240&	$0.71^{+0.39}_{-0.30}$ &	$31^{+49}_{-18} $ \\
\hline
\end{tabular}\\
$^{*}$ Maximal distance $R$ correlates with $R_{h}$: $R=1.41\cdot R_{h}+266$, the correlation coefficient is 0.87 for triplets, and $R=R_{h}$ for pairs.
\end{table*}

An observable agreement with previous results is additional evidence of the correctness of applying the high-order Voronoi tessellation method for identification of pairs and triplets.

\section{Conclusions}

For the first time we introduced and applied the  high-order 3D Voronoi tessellation method for identification of isolated galaxies, pairs and triplets. 

We used volume-limited sample of 6786 galaxies from Northern part of the SDSS DR5 spectroscopic survey (3000 km~s$^{-1}$ $\leq V_{LG} \leq$ 9500 km~s$^{-1}$).  We select single galaxies and pairs by the second-order Voronoi tessellation, as well as triplets by  the third-order Voronoi tessellation method.  As a result we formed 2196 geometric pairs, 1182 triplets and 2394 single galaxies. Then we introduced parameters $p$, $t$ and $s$ to characterize the isolation degree of pairs, triplets and single galaxies, respectively.  We did not do a clear division between physical gravitationally bound systems and non-physical ones, following supposition, that than system is more isolated, the more probability that it is physical. We named the quaters of our single galaxies, pairs and triplets with $s < Q_{1}$,  $p > Q_{3}$, $t > Q_{3}$ , respectively, as "isolated" (i.e. with highest isolation degree). The values of quartiles see in Table~\ref{tab1d}. So, we consider the properties of single galaxies, pairs and triplets in dependence of their isolation degree (in different environment). 

Our main conclusions.

Compact pairs ($R_{h} <$ 150 kpc) and triplets ($R_{h} <$ 200 kpc)  are more isolated in  average than systems in geometric samples, thus they are characterized by different isolation degree. The wider pair (triplet), the smaller isolation degree is observed. Small values of parameters $p_{e}$ and $t_{e}$ are the evidence of loose environment of these systems (they have not a "free space" to be the isolated groups). Thus, we consider wide geometric pairs and triplets as accidental ones in the common field.  

We compared the luminosities of single galaxies and galaxies in geometric pairs and triplets. It was shown that galaxies in isolated pairs and triplets are two times more luminous than isolated galaxies. On one hand it is the evidence of our geometric method accuracy. On the other hand, we can conclude that in such small groups as pairs and triplets the luminosity-density relation is observed.

We considered dark matter content in our groups. The median values of $M_{vir}/L$ for our samples limited by different criteria are 12 $M_{\odot}/L_{\odot}$ for isolated pairs, 44 $M_{\odot}/L_{\odot}$ for isolated triplets, 7 (8) $M_{\odot}/L_{\odot}$ for most compact pairs (triplets) with $R <$ 50 (100) kpc. Note, that for most compact (close or interacting) systems is not very large difference in dark matter content for pairs and triplets, but for isolated triplets the $M_{vir}/L$ is larger in 3 times than for pairs. These results are in agreement with works of other authors. We also found that the pair/triplet is less isolated system (in denser environment), when $M_{vir}/L$ greater.  This relation testifies that galaxy systems in denser environment have greater rms velocity (because of $M_{vir} \sim  S_{v}^{2}$ at fixed distance between galaxies). The $p, t$- $M_{vir}/L$ dependences are observed only for compact systems (up to 150-200 kpc for pairs and up to 250-300 kpc for triplets). 
 
We concluded that 3D Voronoi high-order tessellation method is the effective tool for small groups identification and studying their properties in dependence on environment. In the next paper we will present results on morphological content and colour indices of galaxies in pairs and triplets in comparison with isolated galaxies and their environment.

\section*{Acknowledgments} This work was partially supported by the Cosmomicrophysics Program of the NAS of Ukraine. We are also grateful to Ukrainian Virtual Roentgen and Gamma-Ray Observatory VIRGO.UA and computing cluster of Bogolyubov Institute for Theoretical Physics, for using their computing resources.

\section*{APPENDIX: Radial and projected distance}
To avoid of fingers-of-god effect we used the correction factor in calculation of space distance between galaxies. If the projected $r$ and radial $v$ distances were an equivalent, we should calculate a space distance between two galaxies as
\begin{equation}
d^{2 }=v^{2 }+r^{2 }.
\end{equation}
Obviously we work in the space of radial velocities and under $r$ we mean $r\cdot H_{0}$. We inserted the certain factor $k<1$ as a weight of the radial component. This factor is responsible for the relative virial motion of galaxies. In that case the modified distance is $m^ {2 }=v^ {2 }k^ {2 }+r^ {2 }$.
After some transformations we obtained the equation of ellipse:
\begin{equation}
v^ {2 } \frac {k^ {2 }}{m^ {2 }}+r^ {2 } \frac {1 }{m^ {2 }}=1,
\end{equation}
where $m $ is the minor semiaxis and $ \frac {m }{k }$ is the major semiaxis of the ellipse.
We took into account some tolerance $v_{p}$ in measuring $v$. For that we labelled a major semiaxis as $ \frac {m }{k }=d+v_{p }$, where $d^ {2 }$ is a distance in the space of radial velocities according to (8). So the weight of the radial component is
$k= \frac {m }{d+v_ {p }}$. Since new distance $m$ is a part of formula for $k$, therefore for the simplification we used here $m=d$ and obtained new distance
\begin{equation}
m^ {2 }= \frac {v^ {2 }}{ \left (1+ \frac {v_ {p }}{d } \right )^ {2 }}+r^{2 }.
\end{equation}
In case of $v_ {p } = 0$ the formula (10) changes to (8). If the galaxies are located at the great distance $d >> v_ {p}$, formula (10) also changes to (8). So in these cases the virial velocities have a week action on the distance measurement accuracy. If $d \sim v_ {p }$, than $1+ \frac {v_ {p }}{d }>1, $ i.e. the weight of radial distance decreases. In case $d<<v_{p }$ denominator $\left( 1+\frac{v_{p}}{d}\right)^{2}$ tends to infinity and the radial component loses significance, while projected distance $r$ keeps it. For the our calculations we used $v_ {p }$ = 300 km~s$^{-1}$ as the value of typical relative velocities in small galaxy systems (Ceccarelli et al. 2005). Moreover Karachentsev et al. (1989) showed that majority of physically bound triplets have rms velocity $S_{v} < 300$ km~s$^{-1}$.

\begin{table}
\caption{Physical properties of pairs in dependence on isolation parameter and $v_{p}$.}
\label{tab6}
\begin{tabular}{|c|c|c|c|}
\hline
Pairs &	$N$ &	$S_{v}$ & $R_{h}$ \\
\hline
\multicolumn{4}{|c|}{$v_{p}=$100~km~s$^{-1}$} \\ 
\hline
All geometric & 2211 & $19^{+15}_{-10}$ & $399^{+360}_{-219}$  \\ 	
\hline
$p > 2.93$&	1658 &	$17^{+11}_{-9}$ & $308^{+226}_{-164}$  \\ 	
\hline
$p > 4.72$ &	1005 &	$14^{+8}_{-8}$ & $222^{+163}_{-113} $  \\    	
\hline
$p > 8.15$ & 553 &	$12^{+7}_{-6}$ & $142^{+92}_{-66}$  \\     	
\hline
\multicolumn{4}{|c|}{$v_{p}=$500~km~s$^{-1}$} \\ 
\hline
All geometric & 2210 & $24^{+21}_{-13}$&	$338^{+331}_{-191}$	\\
\hline
$p > 3.83$ & 1657 & $22^{+17}_{-12}$	&$237^{+200}_{-120}$	\\
\hline
$p > 6.15$	& 1105 & $20^{+15}_{-10}$	&$167^{+139}_{-84}	$\\
\hline
$p > 11.19$	& 553 &$17^{+10}_{-9}$	&$100^{+65}_{-45}$	\\
\hline
\end{tabular}
\end{table}

For the testing our method robustness we compared main parameters in cases of different $v_{p}$. Table~\ref{tab6} presents the medians and quartiles of $S_{v}$ and $R_{h}$ for the obtained galaxy pairs at $v_ {p } = 100$~km~s$^{-1}$ and $v_{p } = 500$~km~s$^{-1}$, respectively. The dependences $S_{v}, R_{h}$ on $v_{p}$ are not significant (see Table 2 for comparison). Increasing $v_{p}$ by 5 times produces the changing of rms velocity and projected distance between galaxies by 1.4 times only. In our paper we choose the compromise value $v_ {p }$ = 300~km~s$^{-1}$. So, such approach let us to avoid the mistaken identification of wide pairs with a small velocity difference.

\label{lastpage}


\begin{thebibliography}{}
\bibitem[\protect\citeauthoryear{	Alonso et al. }{ 2004 }]{ alo } Alonso M. S., Tissera B., Coldwell G., Lambas D., 2004, MNRAS, 352, 1081
\bibitem[\protect\citeauthoryear{	Barrena et al.}{2005}]{bar}Barrena R., Ramella M., Boschin W., Nonino M., Biviano A., Mediavilla E., 2005, A\&A,  444,	685
\bibitem[\protect\citeauthoryear{	Barton et al. }{ 2000 }]{ bar1 } Barton E. G., Geller M.J., Kenyon S. J., 2000, ApJ, 530, 660                		
\bibitem[\protect\citeauthoryear{	Barton et al. }{ 2003 }]{ bar } Barton E. G., Geller M.J., Kenyon S. J., 2003, ApJ, 582, 668               		
\bibitem[\protect\citeauthoryear{	Benn \& Wall}{1995}]{ben}Benn C., Wall J., 1995, MNRAS,  272, 678				
\bibitem[\protect\citeauthoryear{	Blanton}{2005}]{bla}Blanton M., Eisenstein D., Hogg D., Schlegel D., Brinkmann J., 2005, ApJ,  629, 143
\bibitem[\protect\citeauthoryear{Ceccarelli}{2005}]{cec}Ceccarelli M., Valotto C., Lambas D., Padilla N., Giovanelli R., Haynes M., 2005, ApJ, 622, 853
\bibitem[\protect\citeauthoryear{	Coles et al.}{1990}]{col}Coles P., Barrow J., 1990, MNRAS,  244, 557				
\bibitem[\protect\citeauthoryear{	Cooper et al.}{2005}]{coo}Cooper M., Newman J., Madgwick D., Gerke B., Yan R., Davis M., 2005, ApJ,  634, 833
\bibitem[\protect\citeauthoryear{	Da Rocha et al.}{2008}]{dar}Da Rocha C., Ziegler B.L., Mendes de Oliveira, C.  2008, MNRAS, 388, 1433
\bibitem[\protect\citeauthoryear{	Propris et al. }{ 2007 }]{ pro } De Propris R., Conselice Ch., Liske J., Driver S., Patton D., Graham A., Allen P., 2007, ApJ, 666, 212	
\bibitem[\protect\citeauthoryear{	Dressler}{1980}]{dre}Dressler A., 1980, ApJ., 236, 351				
\bibitem[\protect\citeauthoryear{	Ebeling et al.}{1993}]{ebe}Ebeling H., Wiedenmann G., 1993, Phys. Rev., 47, 704				
\bibitem[\protect\citeauthoryear{	Einasto et al. }{ 2003 }]{ ein } Einasto M., Einasto J., Muller V., Heinamaki P., Tucker D. L., A\&A, 2003, 401, 851
\bibitem[\protect\citeauthoryear{	Geller at al. }{ 2006 }]{ gel } Geller M.J., Kenyon S. J., Barton E. G., Jarrett T.H., Kewley L.J., 2006, AJ, 132, 2243
\bibitem[\protect\citeauthoryear{	Gerke}{2006}]{ger}Gerke B. F. et al., 2005, ApJ,  625, 6				
\bibitem[\protect\citeauthoryear{	Girardi et al.}{2002}]{gir2}Girardi M., Manzato P., Mezzetti M., Giuricin G., Limboz F., 2002, ApJ, 569,	720
\bibitem[\protect\citeauthoryear{	Gomez et al.}{2003}]{gom}Gomez P., et al. 2003, ApJ, 584, 210			
\bibitem[\protect\citeauthoryear{	Gonzales et al.}{2000}]{gon}Gonzales J. A., Quevedo H., Salgado M., Sudarsky D., 2000, A\&A, 362, 835	
\bibitem[\protect\citeauthoryear{	Icke  \& van de Weygaert}{1987}]{ick}Icke V., van de Weygaert R., 1987, A\&A, 184, 16				
\bibitem[\protect\citeauthoryear{	Ikeuchi \& Turner}{1991}]{ike}Ikeuchi S., Turner E., 1991, MNRAS, 250, 519				
\bibitem[\protect\citeauthoryear{	Jackson}{1972}]{jac}Jackson J.,1972, MNRAS, 156, 1				
			
\bibitem[\protect\citeauthoryear{	Karachentsev }{ 1966 }]{ kar12 } Karachentsev I., 1966, Afz, 2, 81  				
\bibitem[\protect\citeauthoryear{	Karachentsev }{ 1972 }]{ kar10 } Karachentsev I., 1972, Soob. Sp. Astr. Obs., 7, 1            			
\bibitem[\protect\citeauthoryear{	Karachentsev }{ 1987 }]{ kar11 } Karachentsev I., 1987, Double Galaxies (Moskow: Nauka), in Russian		
\bibitem[\protect\citeauthoryear{	Karachentsev et al.}{1989}]{kar9}Karachentsev I., Karachentseva V., Lebedev V., 1989, Izv. SAO, 27, 67		
\bibitem[\protect\citeauthoryear{      Karachentsev \&  Kasparova}{2005}]{ kar7}  Karachentsev I.D., Kasparova A.V.,  2005, Astronomy Letters, 31, 152
\bibitem[\protect\citeauthoryear{	Karachentsev \&  Makarov}{1996}]{kar6}Karachentsev I., Makarov D., 1996, AJ, 111, 794\bibitem[\protect\citeauthoryear{	Karachentsev }{ 1973 }]{ kar13 } Karachentseva V., 1973, Soob. Sp. Astr. Obs., 8, 3 (KIG)\bibitem[\protect\citeauthoryear{	Karachentseva et al. }{ 1979 }]{ kar14 } Karachentseva V. E., Karachentsev I.D., Sherbanobsky A. L., 1979, Izvestia	SAO,	11, 3
\bibitem[\protect\citeauthoryear{	Karachentseva et al. }{ 2000 }]{ kar15 } Karachentseva V.E., Karachentsev I.D., 2000, Astronomy Reports., 44, 501
\bibitem[\protect\citeauthoryear{Karachentseva et al. }{2005}]{karva}Karachentseva V.E., Melnyk O.V., Vavilova I.B., Makarov D.I., 2005, Kin. Phys. Celest. Bodies, 21, 217
				
\bibitem[\protect\citeauthoryear{	Kauffmann et al.}{2004}]{kau}Kauffmann G., White S., Heckman T., Menard B., Brinchmann J., Charlot S., Tremonti C., Brinkmann J., 2004, MNRAS, 353, 713	
\bibitem[\protect\citeauthoryear{	Kiang}{1966}]{kia}Kiang T., 1966, Zeitschrift fur Astrophysik, 64, 433				
\bibitem[\protect\citeauthoryear{	Kim et al.}{2002}]{kim}Kim R., et al., 2002, AJ, 123, 20				
\bibitem[\protect\citeauthoryear{	Lambas et al. }{ 2003 }]{ lam } Lambas D.S., Tissera P.B., Alonso M.S., Coldwell G., 2003, MNRAS, 346, 1189
\bibitem[\protect\citeauthoryear{	Lewis et al.}{2002}]{lew}Lewis J., et al. 2002, MNRAS, 334, 673			
\bibitem[\protect\citeauthoryear{	Lindenbergh}{2002}]{lin} Lindenbergh R. C., 2002, preprint (math-ph/0210345)			
\bibitem[\protect\citeauthoryear{	Lopes et al.}{2004}]{lop}Lopes P., de Carvalho R., Gal R. et al., 2004, AJ, 128, 1017			
\bibitem[\protect\citeauthoryear{	Makarov  \& Karachentsev }{ 2000 }]{ mak1 } Makarov D. I., Karachentsev I. D., 2000, ASP Confer. Ser., 209, 40
\bibitem[\protect\citeauthoryear{	Makarov \& Karachentsev}{2007}]{mak}Makarov D. I., Karachentsev I. D., 2007, Proceed. Int. Astron. Union, IAU Symp., 244, 370	
\bibitem[\protect\citeauthoryear{	Marinoni et al.}{2002}]{mar}Marinoni C., Davis M., Newman J., Coil A., 2002, ApJ, 580, 122			
\bibitem[\protect\citeauthoryear{	Martinez \& Muriel}{2006}]{mart}Martinez H., Muriel H., 2006, MNRAS, 370, 1003				
\bibitem[\protect\citeauthoryear{	Matsuda \& Shima}{1984}]{mat}Matsuda T., Shima E., 1984, Prog. Theo. Phys., 71, 205				
\bibitem[\protect\citeauthoryear{	Melnyk }{ 2006 }]{ mel2 } Melnyk O. V, 2006, Astronomy Letters, 32, 302           				
\bibitem[\protect\citeauthoryear{	Melnyk et al. }{ 2006 }]{ mel1 } Melnyk O. V, Elyiv A. A., Vavilova I. B., 2006, Kinematika I Fizika Nebesnyh Tel,	22, 283 	
\bibitem[\protect\citeauthoryear{	Melnyk et al. }{ 2006 }]{ mel2 } Melnyk O. V, Vavilova I. B., 2006, Kinematika I Fizika Nebesnyh Tel, 22, 422
\bibitem[\protect\citeauthoryear{	Mulchaey et al.}{2003}]{mul}Mulchaey J.,  Davis D., Mushotzky R., Burstein D., 2003, ApJ. Suppl. Ser., 145, 39	
\bibitem[\protect\citeauthoryear{	Nikolic et al. }{ 2004 }]{ nik } Nikolic B., Cullen H., Alexander P., 2004, MNRAS, 355, 874          		
\bibitem[\protect\citeauthoryear{	Panko  \&  Flin}{2006}]{pan}Panko E., Flin P., 2006, The Journal of Astronomical Data, 12, 1			
\bibitem[\protect\citeauthoryear{	Park et al.}{2007}]{par}Park C., Choi Y., Vogeley S., Gott III J., Blanton M., 2007, ApJ, 658, 898	
\bibitem[\protect\citeauthoryear{	Park et al.}{2008}]{par8}Park C., Gott III R., Yun-Young C., 2008, ApJ, 674, 784 			
\bibitem[\protect\citeauthoryear{	Patton et al. }{ 2000 }]{ pat } Patton D.R., Carlberg, R.G., Marzke R.O., Pritchet C.J., da Costa L.N., Pellegrini P.S. 2000, ApJ, 536, 153               	
\bibitem[\protect\citeauthoryear{	Patton et al. }{ 2002 }]{ pat1 } Patton D.R. et al. 2002, AJ, 565, 208		
\bibitem[\protect\citeauthoryear{	Patton et al. }{ 2005 }]{ pat2 } Patton D.R., Grant J.K., Simard L., 2005, AJ, 130, 2043
\bibitem[\protect\citeauthoryear{      	Prada et al. }{2003}]{pra} Prada F. et al., 2003, ApJ, 598, 260.		
	
\bibitem[\protect\citeauthoryear{	Ramella et al.}{2001}]{ram}Ramella M., Boschin W., Fadda D., Nonino M., 2001, A\&A, 368, 776			
\bibitem[\protect\citeauthoryear{	Reda et al.}{2004}]{red}Reda F. Forbes D., Beasley A., O'Sullivan E., Goudfrooij P., 2004, MNRAS, 354, 851	
\bibitem[\protect\citeauthoryear{	Soares }{ 2007 }]{ soa } Soares D.S.L., 2007, AJ, 134, 71                				
\bibitem[\protect\citeauthoryear{	Stocke et al.}{2004}]{sto}Stocke J., Keeney B., Lewis A., Epps H., Schild R., 2004, AJ, 127, 1336			
\bibitem[\protect\citeauthoryear{	SubbaRao et al.}{1992}]{sub}SubbaRao M., Szalay A., 1992, ApJ, 391, 483				
\bibitem[\protect\citeauthoryear{	SubbaRao et al.}{1994}]{sub4}SubbaRao M., Szalay A., Gulkis S., von Gronefeld P., 1994, ApJ, 420, 474
\bibitem[\protect\citeauthoryear{        Tago et al.}{1994}]{sub4}Tago E., Einasto J., Saar E., Tempel E., Einasto M., Vennik J., Muller V., 2008, A\&A, 479, 927		
\bibitem[\protect\citeauthoryear{	Trofimov \& Chernin}{1995}]{tro}Trofimov A.V. \& Chernin A.D., 1995, Astron. Zhurn., 72, 308		
\bibitem[\protect\citeauthoryear{	van de Weygaert \& Icke}{1989}]{van9}van de Weygaert R., Icke V., 1989, A\&A, 213, 1\bibitem[\protect\citeauthoryear{	van de Weygaert}{1991}]{van1}van de Weygaert R., 1991, MNRAS, 249, 159
\bibitem[\protect\citeauthoryear{	van de Weygaert}{1994}]{van4}van de Weygaert R., 1994, A\&A, 283, 361			
\bibitem[\protect\citeauthoryear{	van de Weygaert \& Schaap}{2007}]{van7}van de Weygaert R., Schaap W., 2007, preprint (astro-ph/0708.1441v1)				
				
\bibitem[\protect\citeauthoryear{	Vavilova et al. }{ 2005 }]{ vav } Vavilova I. B., Karachentseva V.E., Makarov D.I., Melnyk O.V., 2005, Kinematika	I Fizika Nebesnyh Tel, 21, 3             	
\bibitem[\protect\citeauthoryear{	Verley et al.}{2007}]{ver}Verley S., et al. 2007, A\&A, 472, 121			
\bibitem[\protect\citeauthoryear{	Voronoi}{1908}]{vor}Voronoi G., 1908, Reine Angew. Math., 134, 198				
\bibitem[\protect\citeauthoryear{	Weinmann et al.}{2006}]{wei}Weinmann S., van den Bosch F., Yang X., Mo H., 2006, MNRAS, 366, 2
\bibitem[\protect\citeauthoryear{	Williams et al.}{1991}]{wil}Williams B., Peacock J., Heavens A., 1991, MNRAS, 252, 43				
\bibitem[\protect\citeauthoryear{	Woods et al. }{ 2006 }]{ woo } Woods D. F., Geller M.J., Barton E.J., 2006, AJ, 132, 197            		
\bibitem[\protect\citeauthoryear{	Xu et al. }{ 2004 }]{ xu } Xu C.K., Sun Y.C., He X.T., 2004, ApJ, 603, L73         				
\bibitem[\protect\citeauthoryear{	Zaninetti}{2006}]{zan}Zaninetti L., 2006, ChJAA, 6, 387

 

\end{thebibliography}
\end{document}